\documentstyle[12pt,aaspp]{article}

\newcommand{\go}{\gtrsim}
\newcommand{\lo}{\lesssim}
\newcommand{\kms}{{\,{\rm km}\,{\rm s}^{-1}}}

\begin{document}

\title{
Collisions of Main-Sequence Stars and the\\
Formation of Blue Stragglers in Globular Clusters
}

\author{James C.~Lombardi, Jr.\altaffilmark{1,2},
Frederic A.~Rasio\altaffilmark{3} and
Stuart L.~Shapiro\altaffilmark{4}}
\altaffiltext{1}{Center for Radiophysics and Space Research, Cornell
University,
Ithaca, NY 14853. Email: lombardi@spacenet.tn.cornell.edu.}
\altaffiltext{2}{Department of Astronomy, Cornell University.}
\altaffiltext{3}{Department of Physics, MIT 6-201, Cambridge, MA 02139.
Email: rasio@mit.edu.}
\altaffiltext{4}{Center for Astrophysics \& Relativity,
  326 Siena Drive, Ithaca, NY 14850}

\begin{abstract}
We report the results of new SPH calculations of parabolic collisions
between two main-sequence stars in a globular cluster.  Such
collisions are directly relevant to the formation of blue stragglers.
In particular, we consider parent stars of mass $M/M_{TO}=0.2,0.5,0.75$,
and $1$, where $M_{TO}$ is the cluster turnoff mass (typically about
$0.8\,M_\odot$). Low-mass stars (with $M=0.2 M_{TO}$ or $0.5 M_{TO}$) are
assumed to be fully convective and are therefore modeled as
$n=1.5$ polytropes. Stars at the turnoff (with $M=M_{TO}$) are
assumed to be mostly radiative and are modeled as $n=3$ polytropes.
Intermediate-mass stars (with $M=0.75 M_{TO}$) are modeled as composite
polytropes consisting
of a radiative core with polytropic index $n=3$ and a convective envelope with
$n=1.5$.
We focus our study on the question of hydrodynamic mixing of helium
and hydrogen, which plays a crucial role in determining the observable
characteristics of blue stragglers. In all cases we find that
there is negligible hydrodynamic mixing of helium into the outer
envelope of the merger remnant. The amount of hydrogen mixed into the core of
the merger depends
strongly on the entropy profiles of the two colliding stars.  For two
stars with nearly equal masses (and hence entropy profiles) very little
hydrodynamic mixing occurs at all, especially if they are close to the
turnoff point.
This is because the hydrogen-rich material from
both stars maintains, on average, a higher specific entropy than the
helium-rich material. If the two parent stars are close to turnoff,
very little hydrogen is present at the center of the merger remnant
and the main-sequence lifetime of the blue straggler could be very short.
In contrast, during a collision between two stars of
sufficiently different masses (mass ratio $q\lo0.5$),
the hydrogen-rich material originally in the smaller star maintains,
on average, a lower specific
entropy than that of the more massive star and therefore settles
preferentially near the center of the merger remnant.
Through this process, moderately massive blue stragglers
(with masses $M_{TO}\lo M_{BS}\lo 1.5M_{TO}$) can obtain a significant supply
of fresh hydrogen fuel, thereby extending their main-sequence lifetime.
Based on our results we conclude that, in contrast to what has been
done in previous studies, blue stragglers formed by direct stellar
collisions should not necessarily be assumed to have initially homogeneous
composition profiles. However, we also demonstrate that the final merged
configurations produced by our hydrodynamic calculations, although very close
to hydrostatic equilibrium,
are usually far from thermal equilibrium. Therefore, it is possible
that convective or rotationally-induced mixing could occur on a thermal
timescale, as the merger remnant recontracts to the main-sequence.

\end{abstract}


\clearpage
\section{Introduction and Motivation}

Blue stragglers are stars that appear along an extension of the main-sequence
(hereafter MS), beyond the turnoff point in the color-magnitude diagram of
a cluster.
It is generally believed that they are more massive objects (mass
$M_{BS}>M_{TO}$)
formed by the merger of two MS stars (each of mass $<M_{TO}$). Merging
can occur through a physical collision, or following the coalescence of the
two components of a close binary system (Leonard 1989; Livio 1993; Stryker
1993;
Bailyn \& Pinsonneault 1995).  Clear evidence for
binary coalescence has been found in the form of contact binaries among
blue stragglers in the low-density globular clusters NGC 5466 (Mateo et
al.\ 1990) and M71 (Yan \& Mateo 1994), as well as in open clusters
(Kalu{\.z}ny \& Ruci{\'n}ski 1993; Milone \& Latham 1994; Jahn, Kalu{\.z}ny
\& Ruci{\'n}ski 1995). Evidence for stellar collisions
comes from recent detections by HST of large numbers of blue stragglers
concentrated in the cores of some of the densest clusters,
such as M15 (De Marchi \& Paresce 1994; Guhathakurta et al.\ 1995)
and M30 (Yanny et al.\ 1994), and from the apparent
lack of binaries in such dense systems  (Shara et al.\ 1995).
Collisions can happen directly between two single stars only in the cores
of the densest clusters, but even in somewhat lower-density clusters they
can also happen indirectly, during resonant interactions involving primordial
binaries (Sigurdsson, Davies, \& Bolte 1994; Sigurdsson \& Phinney 1995;
Davies \& Benz 1995). Observational evidence for the existence of primordial
binaries in globular clusters is now well established (Hut et al.\ 1992;
Cote et al.\ 1994).

Benz \& Hills (1987) performed the first
three-dimensional calculations of direct collisions between two MS
stars.  An important conclusion of their pioneering study was that stellar
collisions could lead to thorough mixing of the fluid.
The mixing of fresh hydrogen fuel into the core
of the merger would reset the nuclear clock of the blue straggler,
allowing it to remain on the MS for up to $\sim10^9\,$yr after its
formation.
In subsequent work it was generally assumed that blue stragglers
resulting from stellar collisions were nearly homogeneous, therefore starting
their life close to the zero-age MS, but with an
anomalously high helium abundance coming from the (evolved) parent stars.
In contrast, little hydrodynamic mixing would be expected to occur during the
much gentler process of binary coalescence, which could take place on a
stellar evolution time scale rather than on a dynamical time scale
(Mateo et al.\ 1990; Bailyn 1992; but see Rasio \& Shapiro 1995, and Rasio
1995).

On the basis of these ideas, Bailyn (1992) suggested a way of distinguishing
observationally between the two possible formation processes. The helium
abundance in the envelope of a blue straggler, which reflects the degree
of mixing during its formation process, can affect its observed position in
a color-magnitude diagram. Blue stragglers made from collisions
would have a higher helium abundance in their outer layers than those made
from binary mergers, and this would generally make them appear somewhat
brighter
and bluer. The analysis was carried out by  Bailyn \& Pinsonneault (1995)
who performed detailed stellar evolution calculations for blue stragglers
assuming various initial profiles. To represent the collisional case, they
again assumed chemically homogeneous initial
profiles with enhanced helium abundances, calculating the total helium mass
from
the age of the cluster and the masses of the parent stars.

In this paper we re-examine the question of mixing in stellar collisions.
We improve on the previous work of Benz \& Hills (1987)
by adopting more realistic stellar models,
and by performing numerical calculations with increased spatial resolution.
We use the smoothed particle hydrodynamics (SPH) method, with $N=3\times10^4$
particles for most of our calculations (Benz \& Hills 1987 also used SPH, but
with
only $N=1024$ particles).

The colliding stars in our calculations are modeled as polytropes or
composite polytropes (Chandrasekhar 1939;
Rappaport, Verbunt, \& Joss 1983; Ruci{\'n}ski 1988), and we adopt a
simple ideal gas equation of state.  The polytropic index $n$
relates the pressure and density profiles in the star according to $P\propto
\rho^{1+1/n}$.  The adiabatic index $\Gamma_1={5\over 3}$ for an ideal
gas and we write the equation of state $P=A \rho^{\Gamma_1}$.  Here
$A$ is a physical parameter related to the local specific entropy $s$
according to $A\propto \exp((\Gamma_1-1)s/k)$, where $k$ is Boltzmann's
constant.
When $\Gamma_1\neq 1+1/n$ the quantity $A$, and hence the entropy $s$, has a
non-zero gradient.
Benz \& Hills (1987) used $n=1.5$, $\Gamma_1={5\over 3}$ polytropic
models to represent MS stars.  Unfortunately, such models
apply only to very low-mass MS stars with large convective envelopes.
For Population II MS stars, the effective polytropic index (defined in
terms of the degree of central mass concentration) is close to $n=1.5$
only for a mass $M\lo0.4\,M_\odot$ (see Lai, Rasio, \& Shapiro 1994,
Table~3).  The object formed by a merger of two such low-mass stars would
hardly be recognizable as a blue straggler, since it would lie below,
or not far above, the MS turnoff point (typically $M_{TO}\simeq 0.8 M_\odot$)
in a color-magnitude diagram.

Stars near the MS turnoff point have very shallow convective envelopes and are
much better described by $n=3$, $\Gamma_1={5\over 3}$ polytropes
(Eddington's ``standard model'', see, e.g., Clayton 1983).
These stars have a density profile much more centrally concentrated
than that of an $n=1.5$ polytropes, and this fact has important
consequences for the hydrodynamics of collisions.
Population II MS stars with masses in the intermediate range
$0.4\,M_\odot \lo M\lo 0.8\,M_\odot$ can be modeled by composite polytropes
with a polytropic index of $n=3$ for the radiative core and
$n=1.5$ for the convective envelope.

Stars close to the MS turnoff point in a cluster are the most relevant to
consider for stellar collision calculations, for two reasons.
First, as the cluster evolves via two-body relaxation,
the more massive MS stars will tend to concentrate
in the dense cluster core, where the collision rate is highest
(see, e.g., Spitzer 1987). Second, collision rates can be increased
dramatically
by the presence of a significant fraction of primordial binaries in the
cluster,
and the more massive MS stars will preferentially tend to be exchanged into
such a binary, or collide with another star, following a dynamical interaction
between two binaries or between a binary and a single star
(Sigurdsson \& Phinney 1995).

Lai, Rasio \& Shapiro (1993) have calculated collisions between MS stars
modeled by $n=3$, $\Gamma_1=5/3$ polytropes, but they focused on
high-velocity (hyperbolic) collisions more relevant to galactic nuclei
than to globular clusters.
The velocity dispersion of globular cluster stars is typically
$\sim10\kms$, which is much smaller than the escape velocity from the stellar
surface.  For example, a star of mass $M=0.8\,M_\odot$
and radius $R=R_\odot$ has an escape velocity $(2GM/R)^{1/2}=552
\kms$.  For this reason, we consider only parabolic collisions in this
paper, i.e., all initial trajectories are assumed to have zero orbital energy.

Our paper is organized as follows.  In \S 2 we describe our
implementation of the SPH method and the numerical setup of our
calculations.  In \S 3, we present the models used for MS stars,
detailing their assumed structure and chemical composition profiles.
We also describe the initial ($t=0$) configuration of the trajectory.
Our results are presented in \S 4. After describing the results for two
typical collisions in detail, we then characterize the rotation states
and the final profiles of all our merger remnants.  We also present a
general method, which does not depend on our particular choice of
initial chemical composition profiles, for calculating the final
profile of any passively advected quantity in the merger remnant.  We
conclude our results with an analysis of the numerical accuracy of our
simulations.  Finally, in \S 5, we discuss the astrophysical
implications of our results as well as directions for future work.

\clearpage

\section{Numerical Method and Conventions}

\subsection{The SPH Code}

Our numerical calculations are done using the smoothed particle
hydrodynamics (SPH) method (see Monaghan 1992 for a recent review).  We
used a modified version of the code developed by Rasio (1991)
specifically for the study of stellar interactions (see Rasio \&
Shapiro 1995 and references therein).  Since SPH is a Lagrangian method,
in which particles are used to represent fluid elements,
it is ideally suited for the study
of hydrodynamic mixing.  Indeed, with the assumption that the gas remains
fully ionized throughout the dynamical evolution,
chemical abundances are passively advected quantities.
Therefore, the chemical
composition in the final fluid configuration can be determined after
the completion of a calculation simply by noting the original and final
positions of all SPH particles and by assigning particle abundances
according to an initial profile.

Associated with each SPH particle
$i$ is its position ${\bf r}_i$, velocity ${\bf v}_i$, mass $m_i$ and a
purely numerical ``smoothing length'' $h_i$ specifying the local
spatial resolution. An estimate of the fluid density at ${\bf r}_i$ is
calculated from the masses, positions, and smoothing lengths of
neighboring particles as a local weighted average,
\begin{equation}
\rho_i=\sum_j m_j W_{ij},
\end{equation}
where the symmetric weights $W_{ij}=W_{ji}$ are calculated from the
method of Hernquist and Katz (1989), as
\begin{equation}
W_{ij}={1\over2}\left[W(|{\bf r}_i-{\bf r}_j|,h_i)+W(|{\bf r}_i-{\bf
                      r}_j|,h_j)\right].
\end{equation}
Here $W(r,h)$ is an interpolation kernel, for which we use the second-order
accurate form of Monaghan and Lattanzio (1985),
\begin{equation}
W(r,h)={1\over\pi h^3}\cases{1-{3\over2}\left({r\over h}\right)^2
      +{3\over4}\left({r\over h}\right)^3,& $0\le{r\over h}<1$,\cr
 {1\over4}\left[2-\left({r\over h}\right)\right]^3,& $1\le{r\over h}<2$,\cr
      0,& ${r\over h}\ge2$.\cr}
\end{equation}

In addition to passively advected scalar quantities (such as the hydrogen and
helium mass fractions $X_i$ and $Y_i$)
each particle~$i$ also carries the local entropy variable
$A_i$.  The specific entropy $s_i$ at ${\bf r}_i$ is
related to $A_i$ by
\begin{equation}
s_i-s_o={k\over \Gamma_1-1} \ln A_i,
\end{equation}
where $k$ is Boltzmann's constant and $s_o$ is a fiducial constant.  Neglecting
radiation pressure, the
pressure at ${\bf r}_i$ is estimated as
\begin{equation}
  p_i=A_i\,\rho_i^{\Gamma_1},
\end{equation}
where $\Gamma_1={5 \over 3}$ is the ratio of specific heats for a fully
ionized ideal gas.  As a self-consistency test, we check that
throughout the dynamical evolution the vast majority of particles have
$p_i$ remaining much larger than the radiation pressure ${1\over
3}aT_i^4$, where $a$ is the radiation constant and $T_i$ is the local
temperature (approximated by assuming an ideal gas).

An SPH code must solve the equations of motion of a large number $N$ of
Lagrangian fluid particles.  Particle positions are updated according
to
\begin{equation}
\dot{\bf r}_i = {\bf v}_i.
\end{equation}
The velocity of particle $i$ is updated according to
\begin{equation}
           m_i \dot{\bf v}_i = {\bf F}^{(Grav)}_i+{\bf F}^{(SPH)}_i
\end{equation}
where ${\bf F}^{(Grav)}_i$ is the gravitational force calculated by a
particle-mesh convolution algorithm (Hockney and Eastwood  1988, Wells
{\it et al.} 1990) based on Fast Fourier Transforms (FFT) on a $128^3$
grid, and
\begin{equation}
{\bf F}^{(SPH)}_i=-\sum_j m_im_j \left[\left({p_i\over\rho_i^2}+
    {p_j\over\rho_j^2}\right)+\Pi_{ij}\right]\nabla_i W_{ij}.
\end{equation}
Here $\Pi_{ij}$ is an artificial viscosity term, while the rest of
equation (8) represents one of many possible SPH-estimators for the
local pressure-gradient force $-m_i(\nabla p/\rho)_i$ (see, e.g., Monaghan
1985).

For the artificial viscosity, a symmetrized version of the form
proposed by Monaghan (1989) is adopted,
\begin{equation}
\Pi_{ij}={-\alpha\mu_{ij}c_{ij}+\beta\mu_{ij}^2\over\rho_{ij}},
\end{equation}
where $\alpha$ and $\beta$ are constant parameters,
$c_{ij}=(c_i+c_j)/2$, and
\begin{equation}
\mu_{ij}=\cases{ {({\bf v}_i-{\bf v}_j)\cdot({\bf r}_i-{\bf r}_j)\over
h_{ij}(|{\bf r}_i
-{\bf r}_j|^2/h_{ij}^2+\eta^2)},& when $({\bf v}_i-{\bf v}_j)\cdot({\bf
r}_i-{\bf r}_j)<0$,\cr
             0,& when $({\bf v}_i-{\bf v}_j)\cdot({\bf r}_i-{\bf
r}_j)\ge0$,\cr}
\end{equation}
with $h_{ij}=(h_i+h_j)/2$.  We have used $\alpha=1$, $\beta=2$ and
$\eta^2=0.01$, which provides a good description of shocks (Monaghan
1989, Hernquist and Katz 1989).

To complete the evolution equations of the fluid, $A_i$ is evolved
according to a discretized version of the first law of thermodynamics:
\begin{equation}
{dA_i\over dt}={\gamma-1\over 2\rho_i^{\gamma-1}}\,
     \sum_jm_j\,\Pi_{ij}\,\,({\bf v}_i-{\bf v}_j)\cdot\nabla_i W_{ij}.
\end{equation}
Equation (11) has the advantage that the total entropy is strictly
conserved in the absence of shocks (ie. when $\Pi_{ij}=0$), and the
disadvantage that the total energy is only approximately conserved
(Rasio 1991; Hernquist 1993).  Both total energy and angular momentum
conservation are monitored throughout the integrations as a measure of
numerical accuracy, and these quantities are conserved typically at the
percent level.

The dynamical equations are integrated using a second-order explicit
leap-frog scheme. Such a low order scheme is appropriate because
pressure gradient forces are subject to numerical noise.  We calculate
the timestep as $\Delta t=C_N\,{\rm Min}(\Delta t_1,\Delta t_2)$ where
$\Delta t_1={\rm Min}_i\,(h_i/\dot{\bf v}_i)^{1/2}$, $\Delta t_2={\rm
Min}_i(h_i/ (c_i^2+v_i^2)^{1/2})$ and the Courant number $C_N=0.8$.
Other details of our implementation, as well as a number of test-bed
calculations using our SPH code are presented in Lombardi, Rasio,
\& Shapiro (1995b).

Twenty of the twenty-three calculations reported here employ
$N=3\times10^4$ equal-mass particles, while the remaining three
calculations (cases U, V and W in Table~2 below) use $N=1.8\times 10^4$
equal-mass particles.  Unequal-mass SPH particles, sometimes used to
allow for higher resolution in low density regions, tend to settle
spuriously to preferred regions in the gravitational potential due to
numerical discreteness effects.  Therefore, calculations with
equal-mass particles are best suited for studying mixing.  In all
cases, time-dependent, individual particle smoothing lengths $h_i$
insure that the spatial resolution remains acceptable throughout the
dynamical evolution and that each particle interacts with a constant
number of neighbors $N_N\simeq64$.  With these resources, the numerical
integration of the SPH equations typically takes about 2 CPU hours per
time unit (eq.~[12]) on an IBM~SP-2 supercomputer.

\subsection{Choice of Units}

Throughout this paper, numerical results are given in units where
$G=M_{TO}=R_{TO}=1$, where $G$ is Newton's gravitational constant and
$M_{TO}$ and $R_{TO}$ are the mass and radius of a terminal-age MS
(TAMS) star at the cluster turnoff point.  The units of time, velocity, and
density are then
\begin{eqnarray}
t_u & = & \left({R_{TO}^3\over G M_{TO}}\right)^{1/2} = 1782\,{\rm s}\times
\left({M_{TO}\over 0.8M_\odot}\right)^{-1/2}\left({R_{TO} \over
R_\odot}\right)^{3/2} \\
v_u & = & \left({G M_{TO}\over R_{TO}}\right)^{1/2} = 391\,{\rm km}\,{\rm
s}^{-1} \times \left({M_{TO}\over 0.8M_\odot}\right)^{1/2}\left({R_{TO}\over
R_\odot}\right)^{-1/2} \\
\rho_u & = & {M_{TO}\over R_{TO}^3} = 5.90\,{\rm g}\,{\rm cm}^{-3}\times
\left({M_{TO}\over0.8M_\odot}\right)\left({R_{TO}\over R_\odot}\right)^{-3}.
\end{eqnarray}
Furthermore, the units of temperature and specific entropy are chosen to be
\begin{eqnarray}
T_u & = & {G M_{TO} m_H \over k R_{TO}} = 1.85\times 10^7\,{\rm K} \times
\left({M_{TO} \over 0.8M_\odot}\right) \left({R_{TO} \over R_\odot}\right)^{-1}
\\
s_u & = & {k \over M_{TO}} = 8.68 \times 10^{-50}\,{\rm erg}\,{\rm
K}^{-1}\,{\rm g}^{-1}\times \left({M_{TO}\over 0.8M_\odot}\right)^{-1}
\end{eqnarray}
where $m_H$ is the mass of hydrogen and $k$ is Boltzmann's constant.

\subsection{Determination of the Bound Mass and Termination of the Calculation}

The iterative procedure used to determine the total amount of
gravitationally bound mass $M$ of a merger remnant is the same as in
Rasio (1991).  Namely, particles with negative specific
enthalpy with respect to the bound fluid's center of mass are
considered bound.  During all of the stellar collisions we considered,
only a small fraction (typically a few percent)
 of the mass is ejected and becomes gravitationally unbound.
Some SPH particles, although bound, are ejected so
far away from the system center of mass that it would take many
dynamical times for them to rain back onto the central remnant and settle
into equilibrium.  Rather than wait for those particles (which would
allow for more spurious diffusion in the central region,
see \S 4.5), we terminate the calculation once we are confident that
at least the inner $95\%$ of the mass has settled into equilibrium.  We
confirm this by two stability tests.  First, we check that the
specific entropy $s$ increases from the center to the surface of the
merger remnant, a sufficient (and necessary for non-rotating stars)
condition for convective stability (see the discussion surrounding
eq.~[18]).  For rotating merger remnants we also check another
dynamical stability criteria, namely that the specific angular momentum
increases from the poles to the equator along surfaces of constant
entropy (Tassoul 1978).

\section{Initial Data}

We consider parent MS stars of masses $M=0.2, 0.5, 0.75$ and $1 M_{TO}$.
The stellar radii are taken from the results of evolution calculations
for Population II stars by D'Antona (1987).
Table~1 lists the values we adopt, as well as the radii enclosing
mass fractions $0.9$ and $0.95$.
For the $0.5 M_{TO}$ star, we adopt the value $0.29 R_{TO}$ for the radius
of the interface between the radiative and convective zones.
These values correspond to an interpolation of the
results of D'Antona (1987) for an age $t=15\,$Gyr.

The $M=0.2$ and $0.5 M_{TO}$ stars are modeled as $n=1.5$ polytropes,
whereas the $M=M_{TO}$ stars are $n=3$ polytropes.  The $0.75 M_{TO}$
stars are modeled as composite polytropes consisting of a radiative
core with index $n=3$ and a convective envelope with $n=1.5$.  Figure~1
shows the specific entropy profiles of these models.  The convective
regions have constant specific entropy. Note that the
specific entropy in the $M=0.2$ and $0.5M_{TO}$ stars is
everywhere smaller than the minimum specific entropy in the two more
massive stars. This fact plays a central role in understanding the dynamics
of the merger involving either an $M=0.2$ or $0.5M_{TO}$ star with a
more massive star.

We have used the stellar evolution code developed by Sienkiewicz and
collaborators (cf.\ Sienkiewicz, Bahcall, \& Paczy\'nski 1990) to
compute the chemical composition profile in the radiative zones of the
$M=0.75$ and $1 M_{TO}$ parent stellar models.  We evolved MS stars of
total mass $M=0.6$ and $0.8\,M_\odot$, primordial helium abundance
$Y=0.25$, and metallicity $Z=0.001$ for a time $t\simeq15\,$Gyr.  This
brought the $M=0.8\,M_\odot$ star to the point of hydrogen exhaustion
at the center.  In the convective regions of our $M=0.5, 0.75$ and $1
M_{TO}$ parent stars, we set a constant helium abundance $Y=0.25$.  For
the $M=0.2 M_{TO}$ star, we set $Y=0.24$ everywhere.  Figure~2 shows
the resulting profiles, which are used to assign the helium abundance to all
the SPH particles in the calculations.  The final column in Table~1
gives the total mass fraction of helium in each of the parent stars.
Although the composition profiles do not affect the hydrodynamics of
a collision in any way, they are needed to determine the chemical composition
profile of the merger remnant.  In \S 4.4, we present a method for
applying our results to arbitrary initial composition profiles.

The stars are initially non-rotating and separated by at least 4 times
the radius of the larger star, which allows tidal effects to be
neglected in the initial configuration.  The initial velocities are
calculated by approximating the stars as point masses on an orbit with
zero orbital energy and a pericenter separation $r_p$.  The Cartesian
coordinate system is chosen such that these hypothetical point masses
of mass $M_1$ and $M_2$ would reach pericenter at positions
$x_i=(-1)^{i}(1-M_i/(M_1+M_2))r_p$, $y_i=z_i=0$, where $i=1,2$ and
$i=1$ refers to the more massive star.  The orbital plane
is chosen to be $z=0$.  With these choices, the center of mass
resides at the the origin.

\clearpage

\section{Results}

Table~2 lists the values of the most important initial parameters and
final results for all the calculations we performed.
The first column gives the label by which the calculation is
referred to in this paper. The second and third columns give the
masses $M_1$ and $M_2$ of the colliding stars, in units of
$M_{TO}\simeq0.8\,M_\odot$. Column~4 gives the ratio $r_p/(R_1+R_2)$,
where $r_p$ is the pericenter separation for the initial orbit and
$R_1+R_2$ is the sum of the two (unperturbed) stellar radii. This ratio
has the value $0$ for a head-on collision, and $1$ for a grazing encounter.
Note, however, that an encounter with $r_p/(R_1+R_2)\go1$ can still
lead to a direct collision in the outer envelopes of the two stars
because of the large tidal deformations near pericenter. We did not
attempt to perform any calculations for $r_p/(R_1+R_2)>1$ here, for
reasons discussed in \S 5.
Column~5 gives the initial separation $r_0$ in units of $R_{TO}$.
Column~6 gives the final time $t_f$ at which the calculation was terminated,
in the unit of equation~(12); see \S 2.3 for a discussion of how the values
of $t_f$ were obtained.
Column~7 gives the number $n_p$ of successive pericenter interactions that
the stars experience before merging. In general, $n_p$ increases with $r_p$,
and it is only for very nearly head-on collisions that the two stars merge
immediately after the first impact ($n_p=1$ in that case).
Column~8 gives the mass-loss fraction $1-M/(M_1+M_2)$, where $M$
is the mass of the bound fluid in the final merged configuration.
Column~9 gives the ratio $T/|W|$
of rotational kinetic energy to gravitational binding energy of the (bound)
merger remnant in its center-of-mass frame at time $t_f$.
Columns~10 and~11 give the velocity
components $V_x$ and $V_y$ in the units of equation (13) for the merger
remnant's center of mass at
time $t_f$ in the system's center-of-mass frame.
Since the amount of mass ejected during a parabolic collision is
very small, the merger remnant
never acquires a large recoil velocity.  The largest
value of 0.035 in our calculations occurs for case~M
and corresponds to a physical speed of about
$14\kms$ (for $M_{TO}=0.8\,M_\odot$ and $R_{TO}=R_\odot$).
This may be large enough to eject the object from the cluster core,
but not to eject it from the entire cluster.

\subsection{Discussion of the Results for Two Typical Cases}

One of our calculations involving two TAMS stars (Case C) has already
been described by Lombardi, Rasio \& Shapiro (1995a).
In this section we discuss in some detail the results of two other
representative cases (E and G).

Figure~3 illustrates the dynamical evolution for Case~E:
a TAMS star ($M_1=M_{TO}$) collides with a slightly less massive star
($M_2=0.75\,M_{TO}$). The initial separation is $r_0=5 R_{TO}$ and the
parabolic trajectory has a pericenter separation $r_p=0.25 (R_1+R_2)$.
The first collision at time $t\simeq 4$ disrupts the outer layers
of the two stars,
but leaves their inner cores essentially undisturbed.  The two
components withdraw to apocenter at $t\simeq 7$, and by $t\simeq 10$
are colliding for the second, and final, time ($n_p=2$).
The merger remnant undergoes some large-amplitude oscillations which
damp away quickly due to shock dissipation.
The final ($t=41$) equilibrium configuration (see
Figure~4) is an axisymmetric, rapidly rotating object ($T/|W|=0.07$).
Figure~5 shows SPH-particle values of the angular velocity $\Omega$ as
a function of radius $r$ in the equatorial plane. We see clearly that
the large envelope of the merger remnant is differentially rotating. The
uniformly rotating core contains only about 15\% of the mass.
The angular velocity drops to half its central value
near $r=1.1 R_{TO}$, and 80\% of mass is enclosed
within the isodensity surface with this equatorial radius.  Only about
2\% of the total mass is ejected during this collision, and the ejection
is nearly isotropic. As a result, the final recoil
velocity of the merger remnant in the orbital plane is only about $0.007$.

Figure~6 displays the thermal energy $U$, kinetic energy $T$,
gravitational potential energy $W$ and total energy $E=U+T+W$ as a
function of time $t$ for case E.  The total energy is conserved to within
$2\%$. Dips in the potential energy $W$ correspond either to
a collision of the two components before final merging or to a maximum
contraction during the subsequent oscillations of the merger remnant.
The criterion we use to distinguish
collisions (which should be included in the number of interactions
$n_p$ before the stars merge) from oscillations is that the first local
maximum of $W$ which is lower than the previous local maximum occurs
immediately after the final merging.
The idea behind this criterion is that a collision without merger
ultimately tends to increase the system's gravitational potential
energy, whereas a merger will decrease the potential energy.  For
example, in Figure~6, the local maximum of $W$ at $t\simeq 11$ is
lower than the one at $t\simeq 6$, so that the dips in $W$ at $t\simeq
4$ and $11$ account for the number $n_p=2$ of interactions given in Table~2 for
case~E.  The remaining dips at $t\simeq 12$ and $15$ correspond to the
peak contraction of the merger remnant during oscillations.  The value
$n_p=2$ obtained here in this way agrees with what one gets
simply by direct visual inspection of the system at various times.
In some cases, however, visual inspection can be
subjective since it is often difficult to recognize two
components connected by a bridge of high-density material
just prior to final merging.

Figure~7 illustrates the dynamical evolution for case~G, which involves
a TAMS star ($M_1=M_{TO}$) and a low-mass MS star with
$M_2=0.5\,M_{TO}$ on a head-on parabolic trajectory with initial
separation $r_0=5$.  The initial collision occurs at time $t\simeq 4$,
and the stars never separate again.  The resulting isodensity surfaces of
the final equilibrium configuration are essentially spherically
symmetric (Figure 8).  About 6\% of the total mass becomes
gravitationally unbound following the collision, and it is ejected
preferentially in the $+x$-direction.  Of this ejected material, 95\%
originated in the more massive ($M=M_{TO}$) star.

Figures~9(a) and~(b) show the entropy profiles for the final
configurations in cases E and G.  Except over the outer few percent of
the mass, where equilibrium has not yet been reached (see \S
2.3), the specific entropy $s$ is an increasing function of the
interior mass fraction $m/M$.  Here $m$ is
the mass inside an isodensity surface, and $M$ is the total bound mass
of the merger remnant.  The scatter of the points in Figure~9 is real, since
isodensity surfaces and surfaces of constant entropy do not coincide.  The
especially small scatter in Figure 9(b)
demonstrates that the entropy does tend towards spherical symmetry in
non-rotating merger remnants, despite the strong angular dependence of
the shock-heating due to the geometry of the collision.

Even though the density and entropy profiles of both the merger remnant
and parent stars are spherically symmetric in case G, this does not
imply that the chemical composition must also share this symmetry.
Indeed, the effects of anisotropic shock-heating are always evident in
the final spatial distribution of the chemical composition.
On a constant-entropy surface in the final configuration,
particles which have been
shock heated the most necessarily had the lowest entropy prior to the
collision.  Since lower entropy material generally has higher helium
abundance (see Figures~1 and 2), shock-heated regions tend to have
higher helium abundances.  Generally, fluid elements which reside in the
orbital plane, and especially those which lie along the collision axis in
$r_p=0$ cases, are
shielded the least from the shock.
Figure~10
displays the angular distribution of the helium
abundance for the merger remnants of cases~E and~G, near the interior
mass fractions $m/M=0.25, 0.5$ and $0.75$.
The helium abundance
$Y$ peaks in Figure~10(b) and (d) at the polar angle $\theta={\pi\over
2}$ (the equatorial plane), as well as in Figure 10(c) at $\phi=0$ (the
collision axis).
In the $r_p\neq 0$ cases, shear in the
differentially rotating merger remnant tends to make the profiles
axisymmetric (see Fig.~10(a)).  However, no dynamical motions exist
to circulate the fluid along the meridional directions, and
consequently, on an isodensity surface, the fractional helium abundance
increases from the poles to the equators for both rotating and non-rotating
merger
remnants (cf. Fig.~10(b and d)).  Meridional circulation will smooth out these
deviations from compositional spherical symmetry over a timescale much longer
than that treatable by our purely dynamical
code (see related discussion in \S 5).  As a
practical concern, we note that stellar evolution codes, which can use our
merger remnants as initial data, usually assume
spherical symmetry.  For these reasons, we often average out the
angular dependence when presenting composition, and other, profiles.

Figures~11(a) and~(b) show the helium mass fraction $Y$ as a function of
the interior mass fraction $m/M$ for the final merged configuration in
cases~E and~G, respectively.  The points correspond to the final SPH
particle values, with the long-dashed curve representing their
average.  The spread in the points is due to the mixing of the fluid as
well as the fact that the final profiles are not spherically
symmetric.  Only a small amount of the observed mixing is due to the
spurious diffusion of SPH particles, i.e., diffusion which is purely a
numerical artifact of the SPH scheme (see \S 4.5).  In case~E
(Figure~11(a)), there is a small amount of hydrogen in the core, with the
innermost 1\% of the mass being 85\% helium and the inner 25\% being
60\% helium.  For both cases~E and~G, it is immediately apparent that
the helium enrichment in the outer layers is minimal since the fractional
helium abundance is just barely above $Y=0.25$, the value in the outer
layers of the parent stars.

The horizontal line at the bottom of Figure~11(b) corresponds to the
particles in case G which originated in the less massive parent star, star~2,
all of which have a helium abundance $Y=0.25$.  Although these
particles are spread over the entire range $0<m/M<1$ in the merger
remnant, they are found preferentially near the center.  Of all the
particles which originated in star~2, 69\% ultimately end up with
$m/M<0.25$, while only 6\% end in the range $0.75<m/M<1$.  Essentially,
the entire star~2 has sunk to the center of the merger remnant,
displacing the material in
star~1 and leaving only a small amount of shock heated gas in the
remnant's outer envelope.  Consequently, the hydrogen enrichment in the core
is quite pronounced; all of the innermost 22\% of the mass originated
in star~2 and is therefore 75\% hydrogen.  Furthermore, the helium
abundance jumps to a maximum average abundance exceeding $Y=0.7$ near
$m/M=0.3$. The subsequent stellar evolution of an object with such an
atypical chemical abundance profile could be quite peculiar.

\subsection{Rotational Properties of the Merger Remnants}

The collisions with $r_p\neq 0$ result in rapidly, differentially
rotating merger remnants. Rotating fluid configurations with
$T/|W|\go0.14$ are secularly
unstable, and those with $T/|W|\go0.26$ are dynamically unstable
(Chandrasekhar 1969; Shapiro and Teukolsky 1983, Chap. 7).
The final merged configurations
are, by definition, dynamically stable, but they could in principle be
secularly unstable.  Although some of our calculations produce
merger remnants close to
the secular instability limit, none of them exceed it.
However, extrapolation of our results to larger values of $r_p$
suggests that secular instabilities could well develop
in some merger remnants.

Table~3 lists the values of the central angular velocity $\Omega_0$ in the
equatorial plane as well as other quantities characterizing the rotation in
the outer layers of the merger remnants. Specifically, for the two mass
fractions $m/M=0.9$ and $m/M=0.95$, we give the values of
the polar and equatorial radii $r_p$ and $r_e$, the angular velocity
$\Omega$ in the equatorial plane, and the ratio $\Omega^2 r_e/g$ of
centrifugal to gravitational acceleration in the equatorial plane.
We see that $\Omega^2 r_e/g$ can be a significant fraction of unity,
indicating that some configurations are rotating near break-up.
The central angular velocity $\Omega_0$ is typically an order of magnitude
larger than the angular velocity $\Omega$ at $m/M=0.95$.

Figure~12 shows contours of the specific angular momentum $\Omega
\varpi^2$, where $\varpi$ is the cylindrical radius measured from the
rotation axis, in the vertical $(x,z)$ plane for several representative
cases. The outermost bounding curves correspond to
$m/M=0.95$. Clearly, the merger remnants are not barotropic since the condition
$d\Omega/dz=0$ is not satisfied everywhere. The
implications of this result will be discussed in \S 5.

\subsection{Interior Structure of the Merger Remnants}

Figures 13(a)--(g) show the variation of the density $\rho$, relative specific
entropy $s-s_o$, helium fraction $Y$ and temperature $T$ as a function of
$m/M$ for all merger remnants.  The density and entropy profiles are
fundamental in the sense that they do not depend on the assumed initial
helium profiles.  The entropy and helium profiles have been averaged
over isodensity surfaces.  The temperature profile is calculated from
the entropy and helium profiles by setting the pressure $p=\rho k T/\mu$ equal
to
$p=A\rho^{\Gamma_1}$, solving for $T$ and using equation~(4).  Here the mean
molecular weight $\mu$ is given by
\begin{equation}
\mu=m_H (2X +{3\over 4}Y+{1\over 2}Z)^{-1},
\end{equation}
where $m_H$ is the mass of hydrogen and $X$, $Y$ and $Z$ are the
fractional abundances of hydrogen, helium and metals.  For Population~II
stars, $Z\sim10^{-4}-10^{-3}$ and the precise value does not
significantly affect the calculated temperature profiles.

Note the peculiar shapes of some of the temperature and helium profiles
in Figure~13. For example, often the temperature or helium abundance
reaches its maximum value somewhere other than the center of the star.
Although these configurations are very close to hydrostatic equilibrium,
it is clear that they are not in thermal equilibrium (see related discussion
in \S 5). These unusual profiles suggest that we look at the condition for
convective stability more carefully.  For a non-rotating star, this condition
can be written simply as
\begin{equation}
{ds \over dr}>0,
\end{equation}
where $s$ is the local specific entropy (see, eg., Landau \& Lifshitz
1957, \S 4).  When written in terms of temperature and composition
gradients, equation~(18) becomes, for an ideal gas,
\begin{equation}
{1 \over T}{dT \over dr}>{1 \over T}\left({dT \over dr}\right)_{ad}+{1 \over
\mu}
{d\mu \over dr}
\end{equation}
which is known as the Ledoux criterion (Kippenhahn \& Weigert 1990, Chap.~6).
Here the subscript $ad$ denotes that the derivative is to be taken at
constant entropy.  Most of our merger remnants
have composition gradients, and it is in the regions where $d\mu/dr>0$
that equation~(19) requires $dT/dr>0$ for stability. For chemically homogeneous
stars,
the second term on the right-hand side of equation~(19) vanishes, and
the familiar Schwarzschild criterion results.
Although equation~(18) is quite general, it does require slight
modification for rotating stars (see Tassoul 1978, Chap.~7).

Figure 13 (a, d, and f) demonstrate that merger remnants formed from
equal mass parent stars have composition profiles which mimic those of
the parents, as can be seen by comparing the resulting helium profiles
to the corresponding parent profiles in Figure 2.  In Figure 13(f), all
of the merger remnants have $Y=0.25$ for all $m/M$, which is simply
because the fully convective parents stars in these cases had $Y=0.25$
everywhere.

We see from Figure~13(c) that the central specific entropy of the
merger remnants increases with $r_p$, which can explain the
qualitatively different shapes of the corresponding helium abundance
profiles.  This increase occurs because the number of interactions
$n_p$, and hence the level of shock heating in star~2 (the smaller
star), increases with $r_p$.  The shock-heating in the central region
of star~1 is less sensitive to $n_p$, since the outer envelope absorbs
the brunt of the shock.  For case~G (solid line), $n_p=1$ and much of
star~2 is able to maintain a lower specific entropy than the minimum
value in star~1.  Since low-entropy material tends to sink to the
bottom of the gravitational potential well, the merger remnant's core
consists entirely of fluid originally from star~2 and therefore with a
helium abundance $Y=0.25$.  For case~H (long-dashed line), $n_p=2$ and,
although the central fractional helium abundance is still $0.25$, there
is enough shock heating for the fluid at small $m/M$ to be affected by
contributions from both stars.  For case~I (short-dashed line), $n_p=3$
and the additional shock heating is sufficient to prevent most of the
fluid from star~2 from reaching the center of the remnant, which
consequently is not significantly replenished with hydrogen.

Figure 13(g) displays the profiles for merger remnants resulting from
collisions between two stars of masses $M_1=M_{TO}$ and $M_2=0.2 M_{TO}$.
In the head-on case, the less massive star (star~2) plummets so
quickly to the center that there is significant
shock-heating in the core of star~1, where the highest fractional
helium abundance resides.  This
causes the helium-rich material to be spread throughout a larger region of
the merger remnant, and the resulting helium profile is not as sharply
peaked as in cases~V and~W.

\clearpage

\subsection{Calculating Final Profiles}

In order to keep our results useful for future applications,
we now present a simple and general method for constructing the
final composition profiles in our calculated merger remnants,
corresponding to any assumed initial composition profiles.
 From our results, we extract some simple functions
which can be applied to transform, for example, any given initial
helium abundance profiles into a final helium abundance profile for
the merger remnant.  Indeed, these transfer functions allow one to find
the final profile of any passively advected quantity,
provided only that the profiles of that quantity in the parent stars are both
known and spherically symmetric.

Table~4 and Figure~14 establish a correlation between the initial and
final mass fractions of a fluid particle.  They also demonstrate that
the details of how the fluid elements are mixed during a collision can
be quite complicated.  Table~4 treats six of the collisions between
equal-mass stars.  The parent stars and merger remnant are partitioned
into zones according to interior mass fractions.  For every zone in the
final configuration, we list the fraction of particles which originated
in each of the initial zones.  Although there is definitely a preferred
final mass fraction $m/M$ for a given initial mass fraction $m_i/M_i$,
there is always a range of $m/M$ obtainable.  In Figure~14, which is
for case G, this range of mass fractions is evident in the spread of
points around a preferred average.  The lower band of points
surrounding the solid line correspond to particles which originated in
star~1, while the upper band surrounding the dashed line correspond to
particles which originated in star~2.  The lines correspond to the
average initial mass fraction $\langle m_i/M_i\rangle$ for stars
$i=1,2$ as a function of the final mass fraction $m/M$, obtained by
binning values of $m/M$.  In contrast, note that if the parent stars
were completely mixed by the collision then the points would be
distributed uniformly over the entire plot with an average initial mass
fraction $\langle m_i/M_i\rangle={1\over 2}$ for all $m/M$.

Let us define $p_1=p_1(m/M)$ to be the probability that a particle with
final mass fraction $m/M$ originated in star~1.  Obviously, $1-p_1$ is
then the probability that the particle originated in star~2.  With
this definition we can approximate the final profile of any passively
advected quantity $Q$ according to
\begin{equation}
Q\left({m\over M}\right)\simeq p_1\left({m\over M}\right)
 Q_1\left(\left.\left\langle{m_1 \over M_1}\right\rangle\right|_{m\over
M}\right)
 +\left(1-p_1\left({m \over M}\right)\right)
 Q_2\left(\left.\left\langle{m_2\over M_2}\right\rangle\right|_{m\over
M}\right),
\end{equation}
where $Q_i$ are the initial (spherically symmetric) profiles for that
quantity in the parent stars $i=1,2$.  The quantities $\left.\langle m_i/M_i
\rangle\right|_{m/M}$ which appear in equation~(20) are the average
initial mass fractions, such as the ones in Figure~14, evaluated at the
final mass fraction $m/M$.  If all particles at $m/M$ came from a
single value of $m_i/M_i$, then equation~(20) would be exact.  In addition,
if the initial profiles are linear over the range of $m_i/M_i$ which
contributes to the abundance at $m/M$, then the above relationship is
exact.

Figures~15(a)--(g) give the average mass fractions $\langle
m_i/M_i\rangle$ as a function of $m/M$ for all of our collisions, while
Figure~16 gives the function $p_1=p_1(m/M)$ for the 12 collisions
involving parent stars of unequal mass. For collisions involving two
identical stars we necessarily have $\langle m_1 /M_1\rangle=\langle
m_2/M_2\rangle$ and $p_1={1 \over 2}$ for all $m/M$.  The solid, long
dashed and short-dashed lines correspond to pericenter separations
$r_p$ of $0, 0.25$, and $0.5(R_1+R_2)$, respectively; in Figure~15(f) the
dot-dashed and dotted lines refer to case S ($r_p=0.75(R_1+R_2)$) and
case T ($r_p=0.95(R_1+R_2)$), respectively.  Note that the horizontal
line $\langle m_i/M_i\rangle={1 \over 2}$ would correspond to the fluid
of star $i$ being completely mixed throughout the merger remnant, which
is not the case for any of our calculations.  For collisions involving
equal-mass stars, if there were no shock heating and no mass loss then
every particle would have identical initial and final mass fractions
(i.e., $m_i/M_i=m/M$) so that the merger remnant's helium profile would
be the same as in the parent stars.  In the equal-mass cases A, B,
C, J, K and L, we do find $\langle m_i/M_i\rangle \simeq m/M$ and the
final helium profiles are indeed quite similar to the parent profile, as shown
in \S 4.3.

Along with equation (20), the functions of Figures~15 and~16 provide
the means for approximating the final profile of any passively advected
quantity.  As a concrete example of how to use this method, we will now
calculate the fractional helium abundance at $m/M=0.28$ in the merger
remnant of case~G, using the same initial profiles as shown in
Figure~2.  From the solid lines corresponding to case G in
Figures~15(c) and 16, we find that$\left.\langle
m_1/M_1\rangle\right|_{0.28}=0.05$, $\left.\langle
m_2/M_2\rangle\right|_{0.28}=0.65$ and $p_1(0.28)=0.62$.  Therefore,
\begin{equation}
Y(0.28)\simeq0.62\times Y_1(0.05)+(1-0.62)\times Y_2(0.65)=0.70,
\end{equation}
where we have used $Y_1(0.05)=0.97$ and $Y_2(0.65)=0.25$, obtained from
the solid and short-dashed lines of Figure~2, respectively.  By
repeating this calculation for other values of $m/M$, we construct the
approximate helium profile shown as the dashed line in Figure~17.  Also
shown for comparison is the ``exact'' profile (solid line) constructed
by considering the individual helium abundance carried by each particle
(the same curve which appears in Figure~11(b)).  We consider the
agreement to be quite good, given the simplicity of the approximation
scheme and the fact that it does not require access to large data files
containing information on all $N=3\times 10^4$ particles.

\clearpage

\subsection{Effects of Spurious Diffusion on Observed Mixing}

In all SPH calculations, numerical noise can lead to spurious, or
artificial, diffusion of SPH particles.  In order to estimate how much
of the observed mixing is in fact caused by particle diffusion, we have
performed a series of systematic tests to evaluate quantitatively the
effects of spurious transport in SPH calculations (Lombardi, Rasio, \&
Shapiro 1995b).  In particular these tests measure, as a function of
the neighbor number $N_N$ and local noise level $v_{rms}$ (the root
mean square particle velocity deviation from the local mean), dimensionless
spurious diffusion coefficients defined by
\begin{equation}
D\equiv{n^{1/3}\over c_s}{d\Delta r_s^2\over dt}
\end{equation}
where $\Delta r_s$ is the distance traveled by a particle due to
spurious diffusion,
$n$ is the local number density of SPH particles, and $c_s$ is
the local sound speed.

Once measured, these diffusion coefficients can be applied to each
particle in our simulations by monitoring its local values of
$v_{rms}$, $n$ and $c_s$ and then estimating how far
that particle has spuriously diffused by numerically integrating
\begin{equation}
\Delta r_s^2=\int{ D {n^{-1/3}\over c_s}dt}.
\end{equation}
{}From the local density gradient at the particle's final position we can
then estimate the equivalent mass fraction corresponding to this
displacement according to
\begin{equation}
\Delta m_s \simeq {\Delta r_s \over 3^{1/2}} |{\bf \nabla} \rho|.
\end{equation}
By repeating this procedure for all the particles, we arrive at an
average spurious diffusion distance $\langle \Delta r_s \rangle$ and
mass fraction  $\langle |\Delta m_s| \rangle$, as well as a
root-mean-square spurious displacement $\langle \Delta r_s^2
\rangle^{1/2}$ and mass fraction $\langle \Delta m_s^2 \rangle^{1/2}$.
This method of estimating spurious diffusion distances will be referred
to as Method~I.

In the case of a head-on collision, another method (which we call Method~II)
can be used, which exploits the axisymmetry around the collision axis
(the x-axis).
If we make the assumption that the entire dynamical evolution remains
axisymmetric (this would not be the case if, e.g., Rayleigh-Taylor
instabilities were to develop), then
a particle should always remain in
the plane containing the collision axis and the particle's initial
position.  The extent to which a particle diffuses in the direction
perpendicular to this plane provides an estimate of the spurious diffusion
distance.  Assuming the diffused amount is the same in all three directions,
the total
spurious diffusion distance is then estimated simply by multiplying by
$3^{1/2}$.  We finally convert the spurious diffusion distance to
an equivalent mass fraction exactly as in Method~I (see eq.~[24]).

The results of the two methods applied to our calculations are given
in Table~5. When two numbers are given, the second one is
calculated by Method II.  Included in this table are the average
spurious diffusion distance $\langle \Delta r_s \rangle$, the root-mean-square
diffusion distance $\langle \Delta r_s^2\rangle ^{1/2}$, the
average equivalent mass fraction $\langle |\Delta m_s| \rangle/M$,
and the root-mean-square mass fraction $\langle \Delta
m_s^2\rangle ^{1/2}/M$.
It is clear that the two methods are generally in good agreement,
and that, when expressed in terms of $m/M$, the effects of
spurious diffusion are always small.

Also listed in Table~5
are the observed average ($\langle |\Delta m_o| \rangle/M$)
and root-mean-square ($\langle \Delta m_o^2\rangle ^{1/2}/M$) total deviation
in final mass fraction.
The last column then subtracts from the total square deviation
$\langle \Delta m_o^2\rangle$ the contribution  $\langle \Delta m_s^2\rangle$
from spurious diffusion.
For example, in case~G, we observe a root-mean-square change in the
interior mass fraction of ${\langle\Delta m_o^2\rangle^{1/2}/
M}=0.09$ over the entire calculation.  Using Method I, we estimate that
the root-mean-square change in interior mass fraction due to spurious
diffusion is ${\langle\Delta m_s^2\rangle^{1/2}/ M}\simeq 0.036$,
while Method II gives an estimate of $0.057$ for this quantity.  We
therefore believe that the physical root-mean-square spread (i.e., the
spread in a calculation free of spurious diffusion) would be
approximately $0.08$ or $0.07$, depending on whether Method~I
or Method~II is more accurate.

\clearpage

\section{Summary and Discussion}

The main results of this paper can be summarized as follows.  We have
demonstrated that the typical merger remnants produced by collisions
are rapidly and differentially rotating, and are far from chemically
homogeneous, with composition profiles that can be rather peculiar in
certain cases. For example, it often happens that the maximum helium
abundance does not occur at the center of the remnant
(cf.\ Fig.~11(b)).  The merger remnants produced by our dynamical
calculations, although very close to hydrostatic equilibrium, are
usually far from thermal equilibrium, as discussed below.  In
particular, we have shown that the merger remnants are not barotropes
(ie., the condition $d\Omega/dz=0$ is generally not satisfied), and
that their temperature profiles can have positive gradients ($dT/dr>0$)
in certain regions.

At a qualitative level, many of our results can be understood very
simply in terms of the requirement of convective stability of the final
merger remnant.  If entropy production in shocks could be neglected
(which may be reasonable for parabolic collisions, especially in the
head-on case), then one could predict the qualitative features of the
remnant's composition profile simply by observing the composition and
entropy profiles of the parent stars.  Convective stability requires
that the specific entropy $s$ increase from the center to the surface
($ds/dr>0$) in the final hydrostatic equilibrium configuration.
Therefore, in the absence of shock-heating, fluid elements conserve
their entropy and the final composition profile of a merger remnant
could be determined simply by combining mass shells in order of
increasing entropy, from the center to the outside. Many of our results
follow directly.  For example, in the case of a collision between two
identical stars, it is obvious why the composition profile of the
merger remnant remains very similar to that of the parent stars. For
two stars of very different masses, the much lower-entropy material of
the lower-mass star tends to concentrate at the center of the final
configuration, leading to the unusual composition and temperature
profiles seen in Figure~13 (c,~e, and~g).

Regions where the dynamical stability criterion $ds/dr>0$ (eq.~[18]) is
satisfied can nevertheless be thermally, or secularly, unstable. The
small vertical oscillations (at the local Brunt-V\"ais\"al\"a frequency
$\Omega_{BV}\propto [ds/dr]^{1/2}$) of a fluid element in such a region
have amplitudes that grow slowly, and mixing will occur on a timescale
comparable to the local radiative damping time (see, e.g., Kippenhahn
\& Weigert, Chap.~6).  The thermal instability can be of two types.
When $d\mu/dr>0$ and $dT/dr>0$ (as in Figure 13 (c,~e, and~g)), a
so-called thermohaline instability can develop, allowing fingers of
the high-$\mu$ material to penetrate down into the lower-$\mu$, colder
material below (see, e.g., Ulrich 1972).  When such mixing occurs in the
stellar core, it tends
to increase the central helium abundance and therefore decrease the
time that the merger remnant can remain on the MS.

When $d\mu/dr<0$ but $dT/dr<(dT/dr)_{ad}$, so-called semiconvection
occurs (Spruit 1992).  In terms of easily computed SPH variables, this
criterion is equivalent to
\begin{equation} 0<{1 \over A}{dA \over dr}<-{\Gamma_1 \over \mu}{d\mu \over
dr} \end{equation}
where $A$ is related to specific entropy by equation (4).  We have
tested our merger remnants formed from head-on collisions and found that
this instability is typically present.  Figure~18 shows, as a function
of the final mass fraction $m/M$, the fraction $f_{sc}$ of gas which is
semiconvective for six of our merger remnants.  In all cases no
semiconvective instability exists in the outer $\sim$20\% of the mass,
so that we do not expect this mixing mechanism can increase the helium
abundance of the outer layers.  Figure 18 does demonstrate, however,
that some merger remnants (those of cases A, D and J) have a unstable
region which extends to the center, and these remnants therefore have a
means of mixing hydrogen into their cores.  For instance in case A, we
see that the inner $\sim$40\% could be significantly affected.  As
semiconvection slowly attempts to decrease the central helium fraction,
it must compete against hydrogen burning.  In addition, the right hand
side of equation (25) changes as the fluid mixes, so that the details
of this complicated process can only be followed numerically with a
stellar evolution code.

For a rotating, chemically homogeneous star, stable thermal equilibrium
requires $d\Omega/dz=0$, where $\Omega$ is the angular velocity and $z$
is measured along the rotation axis (the Goldreich-Schubert stability
criterion; see, e.g., Tassoul 1978, Chap.~7).
{}From the representative set of specific angular momentum
contours presented in Figure~12, it is therefore evident that the
merger remnants of cases Q,~R,~S and~T (which are chemically
homogeneous, since their parent stars were fully mixed) cannot be in
thermal equilibrium.  In chemically inhomogeneous stars, regions with a
sufficiently large and stabilizing composition gradient ($d\mu/dr<0$)
can in principle still be thermally stable even with $d\Omega/dz \neq
0$.  However, it seems unlikely that the composition profiles generated
dynamically by a collision would conspire to keep the remnants
everywhere thermally stable.

Deupree (1990) has shown that stars with rapidly rotating cores and
slowly rotating envelopes can have their MS lifetime extended beyond
that of their non-rotating counterparts.  The fact that much of the
angular momentum is hidden deep in the remnant's interior suggests a
possible explanation for why observations of blue stragglers in open
clusters such as M67 find no signs of rapid rotation (Peterson, Carney,
\& Latham 1984; Mathys 1991).  Recently, Leonard \& Livio (1995) have
argued that the spin-down timescale of blue stragglers due to magnetic
breaking should be on the order of only $10^5$ years, which is much
less than the thermal timescale of approximately $10^7$ years, so that
initially rapidly rotating merger remnants may not be a problem for the
collisional formation scenario.

Stellar encounters with separations $r_p$ larger than those considered
in this paper are difficult to compute directly with SPH since the
amount of energy $\Delta E$ dissipated during the first interaction is
then so small that the integration time until the next pericenter
passage can be several orders of magnitude larger than the hydrodynamic
time.  We believe that our results can be safely extrapolated all the
way to values of $r_p\simeq1.2(R_1+R_2)$.  For instance, from our
results for collisions of equal-mass stars it seems very likely that
the helium profile of the merger remnant will always mimic that of the
parent stars. In Figure~13(f), note that the density, entropy and
temperature profiles also seem to be converging onto a fixed profile,
and that the profiles for the $r_p=0.5, 0.75$ and $0.95(R_1+R_2)$ cases
all look very similar.  For $r_p\go1.2(R_1+R_2)$ the encounter is
better described as a tidal capture than a collision, i.e., the amount
of energy dissipated is sufficient to form a bound system, but no
direct collision occurs, even in the outer layers of the stars. The
maximum value $r_p=r_{cap}$ for tidal capture can be calculated
accurately from linear perturbation theory (Press \& Teukolsky 1977;
McMillan, McDermott, \& Taam 1987). For two identical $0.8\,M_\odot$ MS
stars and a relative velocity at infinity $v_\infty=10\kms$, McMillan
et al.\ (1987) find $r_{cap}/(R_1+R_2)\simeq1.4$, which leaves little
room for ``clean'' tidal captures. In addition, the long-term evolution
of a tidal-capture binary may well lead to merging of the two stars
even if the initial interaction is in the linear regime (for recent
discussions, see Mardling 1995a,b and Kumar \& Goodman 1995).

It must be stressed that the amount of mixing determined by SPH
calculations is always an upper limit. Indeed, some of the mixing
observed in a calculation will always be a numerical artifact.
Low-resolution SPH calculations in particular tend to be very noisy and
the noise can lead to spurious diffusion of SPH particles, independent
of any real physical mixing of fluid elements. In \S4.5 we introduced
two simple methods to evaluate quantitatively the effects of spurious
diffusion in our calculations.  The results suggest that spurious
diffusion does not significantly corrupt our results.  This is seen for
example in the near agreement of numbers in the last two columns of
Table~5.  However, both methods are approximate.  The diffusion
coefficients used in Method~I have been measured in the absence of
artificial viscosity.  The presence of artificial viscosity in our
collision calculations may slightly change the effective values of
these coefficients.  Method~II assumes that we can neglect any
nonaxisymmetric instabilities in a head-on collision and that spurious
diffusion is isotropic (which may not be true in the presence of strong
entropy gradients).
Nevertheless, the reasonable agreement between the two methods
(cf.\ Table~5) gives us confidence that these simplifying assumptions
are generally satisfied to a good approximation.  The general question
of spurious transport in SPH calculations will be addressed in a
separate paper (Lombardi et al.\ 1995b).

Benz \& Hills (1987) performed the first fully three-dimensional
calculations of collisions between two identical MS stars. In contrast
to the present work, they considered only $n=1.5$ polytropes, which are
mostly relevant for collisions of very low-mass stars ($M_1=M_2\lo
0.4M_\odot$).  Their calculations for parabolic collisions indicated a
higher level of mixing than we find in similar cases (cases P, Q, R, S
and T in our Table~1).  We have also computed a number of collisions
using $N=1024$ SPH particles (as in Benz \& Hills 1987, rather than
$N=3 \times 10^4$ particles as in most of our other calculations), and
found substantially higher levels of spurious diffusion.  We conclude
that a significant part of the mixing observed by Benz \& Hills (1987)
was a numerical artifact.  However, we should note that our
higher-resolution  calculations for $n=1.5$ polytropes (cases P, Q, R,
S and T) do exhibit a generally higher level of mixing than observed
for other models (e.g., cases A, B and C). This is not surprising since
we expect parent stars of constant entropy, which are only marginally
stable against convection, to be easier to mix than those with
significant positive entropy gradients (stable stratifications).  In
addition, the more homogeneous density profile of $n=1.5$ polytropes
leads to a better distribution of the impact energy throughout the
entire mass of fluid.  Benz \& Hills (1992) have performed calculations
of collisions between $n=1.5$ polytropes with a mass ratio
$M_2/M_1=0.2$, using $3500$ SPH particles per star. These calculations
are not directly relevant to blue straggler formation, given the very
low masses of the MS stars involved ($M_1\lo0.4\,M_\odot$, hence
$M_2<0.1\,M_\odot$), and, since none of our calculations model such low
masses, no direct comparison is possible.

There are a number of ways by which our results could be improved or
extended.  For instance, the equation of state could be extended to
include radiation, partial ionization and electron degeneracy
corrections. We do not expect these corrections to be significant.
More accurate initial composition profiles could be also used.
Note, however, that our results can be applied to
arbitrary initial profiles by the method of \S 4.4.
Incidentally, profiles of $^7$Li would be particularly interesting to
consider  since this element is destroyed at
temperatures $T\go 10^6 K$ and is therefore an
observationally measurable indicator of mixing (see, e.g., Hobbs \&
Mathieu 1991 and Pritchet \& Glaspey 1991).

Our dynamical calculations and the determination of hydrodynamical
mixing are only the first step in modeling blue straggler formation.
The merger remnants, which are much larger than normal equilibrium MS
stars of the same mass, will recontract to the MS on a thermal
timescale ($\sim10^7\,$yr).  As they evolve, other mixing processes
such as meridional circulation and convection may well be important.
Calculations of this thermal relaxation phase using the results of
dynamical calculations, such as those presented in this paper, as
initial conditions, will be necessary in order to make detailed
predictions for the observable parameters of blue stragglers.

Recently, Sills, Bailyn and Demarque (1995, hereafter SBD) have begun
investigating the consequences of blue stragglers being born unmixed.
To create their unmixed initial model of a blue straggler formed by the
collision of two TAMS stars, SBD relaxes a non-rotating TAMS star whose
mass has been artificially doubled but which is otherwise unchanged.
The subsequent stellar evolution is contrasted to that of a fully mixed
(ie. chemically homogeneous) blue straggler.  SBD finds that the high
central helium concentration in the unmixed models causes the time
spent on the MS ($\sim5\times10^7$ yr) to be drastically shorter than
for the fully mixed counterparts ($\sim5\times10^8$ yr), making it
difficult to account for the observed numbers blue stragglers in the
core of NGC 6397.  In addition, such unmixed blue stragglers are
neither bright nor blue enough to explain the observations.  A blue
straggler population consisting purely of non-rotating, unmixed merger
remnants of two TAMS parent stars is therefore not sufficient to
explain the core blue stragglers in NGC 6397.

Future work which follows the approach of SBD would clearly be
beneficial.  A number of factors need to be considered in more detail.
For instance, it is unrealistic to expect that all collisional blue
stragglers are born only from TAMS parent stars.  Instead, attention
must also be given to collisions between unequal mass parent stars,
which form merger remnants with profiles that are neither homogeneous
nor like that of the parents.  That is, such remnants are neither fully
mixed nor unmixed.  Since these blue stragglers have an enhanced
hydrogen abundance in their cores, they will presumably remain on the
MS for a longer time and have a different position on a color magnitude
diagram than their unmixed counterparts.  Moreover, most blue
stragglers will be formed rapidly rotating, especially in the stellar
core, which acts to extend their postponed residence on the MS (Deupree
1990).  In addition, the density and specific entropy profiles shown in
Figure~13 can be used to specify the structure of the blue straggler
and improve upon SBD's somewhat {\it ad hoc} profiles obtained by
artificially scaling the mass of an equilibrium star.

\bigskip
\acknowledgments

Support for this work was provided by NSF Grant AST 91--19475,
NASA Grant NAG 5--2809, and NASA Grant HF-1037.01-92A.
Computations were performed at the Cornell Theory Center, which
receives major funding from the NSF and IBM Corporation, with
additional support from the New York State Science and Technology
Foundation and members of the Corporate Research Institute.

\clearpage

% FIG. 1
\begin{figure}
\caption{
Specific entropy $s$, relative to the constant $s_o$,
as a function of radius $r$ for the parent stars used in
our collision calculations.  The dotted, short-dashed, long-dashed and
solid curves correspond to parent stars of mass $M=0.2, 0.5, 0.75$ and
$1M_{TO}$, respectively, where $M_{TO}$ is the mass of a turnoff star.
Units are discussed in \S 2.2.
}
\end{figure}

% FIG. 2
\begin{figure}
\caption{
Fractional helium abundance $Y$ as a function of interior mass fraction
$m/M$ for the parent stars whose entropy profiles are shown in
Figure~1.  As in Figure~1, the dotted, short-dashed, long-dashed and
solid curves correspond to parent stars of mass $M=0.2, 0.5, 0.75$ and
$1M_{TO}$, respectively.
}
\end{figure}

% FIG. 3
\begin{figure}
\caption{
Snapshots of density contours in the orbital plane ($z=0$) showing the
dynamical evolution in case~E, where we consider a parabolic collision
between parent stars of masses $M_1=M_{TO}$ and $M_2=0.75 M_{TO}$ at a
pericenter separation $r_p=0.25 (R_1+R_2)$.  There are 8 density
contours, which are spaced logarithmically and cover 4 decades down
from the maximum.
}
\end{figure}

% FIG. 4
\begin{figure}
\caption{
Density contours and velocity field for the final ($t=41$)
configuration of the case~E collision in (a) the equatorial ($z=0$)
plane and (b) the $y=constant$ plane which includes the rotation axis.
There are 10 contours such that, starting from the center, each
corresponding isodensity surface encompasses an additional 10\% of the
total mass, with the exception of the outermost contour which
encompasses 95\% of the mass.
}
\end{figure}

% FIG. 5
\begin{figure}
\caption{
Angular velocity $\Omega$ as a function of radius $r$ for particles
near the equatorial plane ($|z_i|<2 h_i$, where $h_i$ is the particle
smoothing length) in the final ($t=41$) merger remnant of case~E.
}
\end{figure}

% FIG. 6
\begin{figure}
\caption{
The internal energy $U$, kinetic energy $T$, gravitational potential
energy $W$ and total energy $E=U+T+W$ as a function of time $t$ for
case~E.
}
\end{figure}

% FIG. 7
\begin{figure}
\caption{
Snapshots of density contours in the orbital plane ($z=0$) showing the
dynamical evolution in case~G, where we consider a head-on parabolic
collision between parent stars of masses $M_1=M_{TO}$ and $M_2=0.5
M_{TO}$.  There are 8 density contours, which are spaced
logarithmically and cover 4 decades down from the maximum.
}
\end{figure}

% FIG. 8
\begin{figure}
\caption{
Cross-sections of isodensity surfaces for the final ($t=30$)
configuration of the case~G collision in (a) the equatorial ($z=0$)
plane and (b) the $y=constant$ plane which includes the rotation axis.
The 10 contours are spaced the same as in Figure~4.
}
\end{figure}

% FIG. 9
\begin{figure}
\caption{
Relative specific entropy $s-s_o$ as a function of final mass fraction
$m/M$ for the merger remnants of (a) case E and (b) case G.  Here,
$m$ is the mass enclosed by an isodensity surface, and $M$ is the total
bound mass of the merger remnant.
}
\end{figure}

% FIG. 10
\begin{figure}
\caption{
Fractional helium abundance $Y$ for particles in the vicinity of the
final mass fractions $m/M={1 \over 4}, {1 \over 2}$ and ${3 \over 4}$ as
a function of the azimuthal angle $\phi$ (measured counter-clockwise
from the positive x-direction) and the polar angle $\theta$ (measured
from the rotation axis). Figures~10(a) and~(b) are for the merger
remnant of case~E while Figures~10(c) and~(d) are for case~G.
}
\end{figure}

% FIG. 11
\begin{figure}
\caption{
Fractional helium abundance $Y$ as a function of final mass fraction
$m/M$ for the merger remnants of (a) case~E and (b) case~G.  The dashed
line represents the average of the individual SPH particle values.
}
\end{figure}

% FIG. 12
\begin{figure}
\caption{
Contours of the specific angular momentum $\Omega \varpi^2$ in the
vertical $(x,z)$ plane (meridional section) for several representative
cases. Here $\Omega$ is the angular velocity and $\varpi$ is the
cylindrical radius measured from the rotation axis. The contours have a
linear spacing of $0.1(G M_{TO}R_{TO})^{1/2}$, with the specific
angular momentum increasing from the rotation axis to the outer layer
of the merger remnant.  The thick bounding curve marks the isodensity
surface which encloses 95\% of the total gravitationally bound mass.
}
\end{figure}

% FIG. 13
\begin{figure}
\caption{
Interior profiles of the final merger remnants.  The density $\rho$,
relative specific entropy $s-s_o$, fractional helium abundance $Y$ and
temperature $T$ are shown for cases: (a) A, B \& C, (b) D, E \& F, (c)
G, H \& I, (d) J, K \& L, (e) M, N \& O, (f) P, Q, R, S \& T and (g)
U, V \& W.  The solid, long-dashed and short-dashed lines correspond to
collisions with pericenter separation $r_p=0, 0.25$ and $0.5(R_1+R_2)$,
respectively.  In (f), the dot-dashed and dotted lines represent cases
S and T, which have $r_p=0.75$ and $0.95(R_1+R_2)$, respectively.  The
density and specific entropy profiles are plotted in the same frame
($\rho$ is maximum at $m/M=0$, while $s$ is minimum there).  The units
are discussed in \S 2.2.
}
\end{figure}

% FIG. 14
\begin{figure}
\caption{
Individual points show the initial mass fraction $m_i/M_i$ of SPH
particles which originated in star $i=1$ (the more massive star) or
$i=2$ (the less massive star) as a function of the particles' final
mass fraction $m/M$ for case G.  The lines represent averages $\langle
m_i/M_i \rangle$ obtained by binning in $m/M$, with the solid line
corresponding to star~1 and the dashed line corresponding to star~2.
}
\end{figure}

% FIG. 15
\begin{figure}
\caption{
Average initial mass fractions $\langle m_i/M_i \rangle$ as a function
of final mass fraction $m/M$ for cases: (a) A, B \& C, (b) D, E \& F,
(c) G, H \& I, (d) J, K \& L, (e) M, N \& O, (f) P, Q, R, S \& T and
(g) U, V \& W.  As in Figure~13, the different lines
correspond to different pericenter separations $r_p$ for the initial
orbit.  For collisions involving two identical parent stars, $\langle
m_1/M_1 \rangle=\langle m_2/M_2 \rangle$, so that only one set of plots
is necessary.
}
\end{figure}

%FIG. 16
\begin{figure}
\caption{
Probability $p_1$ that a particle in the merger remnant originated
in star~1, as a function of
final mass fraction $m/M$ for cases: (a) D, E \& F, (b) G, H \&
I, (c) M, N \& O and (d) U, V \& W.
The solid, long-dashed and short-dashed lines
correspond to collisions with pericenter separation $r_p=0$, $0.25$ and
$0.5(R_1+R_2)$, respectively.  Collisions involving stars of equal mass
necessarily have $p_1={1 \over 2}$, and therefore $p_1$ is not displayed
for such cases.
}
\end{figure}

% FIG. 17
\begin{figure}
\caption{
The exact (solid) and approximate (dashed) helium abundance
profiles for the final merger remnant of case~G.  The exact profile was
calculated using the values of $Y_i$ (for all $N=3\times 10^4$ particles)
derived from our assumed initial composition profiles.
The approximate profile was derived by using equation~(20) and
the curves of Figures~15 and~16 corresponding to case~G.
}
\end{figure}

% FIG. 18
\begin{figure}
\caption{
The fraction $f_{sc}$ of gas unstable to semiconvection as a
function of the mass fraction $m/M$ in the merger remnants of the
head-on cases A (solid line), D (long-dashed line), G (short-dashed
line), J (short- and long-dashed line), M (dotted and short-dashed line) \& U
(dotted line).
}
\end{figure}

\end{document}